\documentclass[letter]{aa} 

\usepackage{graphicx}
\usepackage{txfonts}
\usepackage{xcolor}
\usepackage{comment}
\usepackage{arydshln}
\def\msun{{\rm M_{\odot}}}

\def\be{\begin{equation}}
\def\ee{\end{equation}}

\def\del#1{{}}

\newcommand\mj{{\,{\rm M}_{\rm J}}}

\begin{document}

   \title{Accretion bursts in high-mass protostars: A new test bed for models of episodic accretion }


   \author{Vardan G. Elbakyan\inst{1,2}\fnmsep\thanks{vardan.elbakyan@leicester.ac.uk}, Sergei Nayakshin\inst{1}, 
   Eduard I. Vorobyov\inst{3,4,2}, Alessio Caratti o Garatti\inst{5}, Jochen Eislöffel\inst{6}
          }

   \institute{Department of Physics and Astronomy, University of Leicester, Leicester LE1 7RH, UK
          \and
             Research Institute of Physics, Southern Federal University, Rostov-on-Don 344090, Russia
        \and
            University of Vienna, Department of Astrophysics, Vienna, 1180, Austria
        \and
            Ural Federal University, 51 Lenin Str., 620051 Ekaterinburg, Russia
        \and
            Dublin Institute for Advanced Studies, 31 Fitzwilliam Place, D02 XF86, Dublin, Ireland
        \and
            Thüringer Landessternwarte Tautenburg, Sternwarte 5, 07778 Tautenburg, Germany
             }

   \date{Received May 15, 2021; accepted June 16, 2021}
   
   \titlerunning{HMYSO outbursts}
   \authorrunning{Elbakyan et al.}

 
  \abstract
   {}
   {It is well known that low-mass young stellar objects (LMYSOs) gain a significant portion of their final mass through episodes of very rapid accretion, with mass accretion rates up to $\dot M_* \sim 10^{-4} \msun$~yr$^{-1}$. Recent observations of high-mass young stellar objects (HMYSOs) with masses  $M_* \gtrsim 10 \msun$ uncovered outbursts with accretion rates exceeding $\dot M_*\sim 10^{-3}\msun$~yr$^{-1}$. Here, we examine which scenarios proposed in the literature so far to explain accretion bursts of LMYSOs can also apply to the episodic accretion in HMYSOs.}
   {We utilise 1D time-dependent models of protoplanetary discs around HMYSOs to study burst properties.} 
   {We find that discs around HMYSOs are much hotter than those around their low-mass cousins. As a result, a much more extended region of the disc is prone to the thermal hydrogen ionisation and magnetorotational (MRI) activation instabilities. The former, in particular, is found to be ubiquitous in a very wide range of accretion rates and disc viscosity parameters.   The outbursts triggered by these instabilities, however, always have too low of an $\dot M_*$ and are one to several orders of magnitude too long compared to those observed from HMYSOs to date. On the other hand, bursts generated by tidal disruptions of gaseous giant planets formed by the gravitational instability of the protoplanetary discs yield properties commensurate with observations, provided that the clumps are in the post-collapse configuration  with planet radius $R_{\rm p} \gtrsim 10 $ Jupiter radii.  Furthermore, if observed bursts are caused by disc ionisation instabilities, then they should be periodic phenomena with the duration of the quiescent phase comparable to that of the bursts. This may yield potentially observable burst periodicity signatures in the jets, the outer disc, or the surrounding diffuse material of massive HMYSOs. Bursts produced by disruptions of planets or more massive objects are not expected to be periodic phenomena, although multiple bursts per protostar are possible.}   
   {Observations and modelling of episodic accretion bursts across a wide range of young stellar object (YSO) masses is a new promising avenue to break the degeneracy between models of episodic accretion in YSOs.}


   \maketitle
%


\section{Introduction}

FU Orionis-type stars, or FUors, are low-mass young stellar objects (LMYSOs) that show accretion outbursts in which the accretion rate onto the young stellar object (YSO) rises from $\dot M_* \sim (10^{-7} - 10^{-8})~M_{\odot}\rm{yr^{-1}}$ to $(10^{-5} - 10^{-4})~M_{\odot}\rm{yr^{-1}}$ for a few tens to $\gtrsim$100~yrs \citep{2014AudardAbraham}. Observations and modelling suggest that, on average, LMYSOs go through a dozen such outbursts and that these outbursts contribute significantly to the final mass of the star \citep{1996HartmannKenyon,2012DunhamVorobyov}. 

A number of scenarios have been proposed to explain FUors. The disc could undergo thermal instability \citep[TI;][]{1994BellLin, 1996KleyLin}, in which hydrogen is ionised, which increases disc viscosity. However, TI outbursts are insufficiently bright and are also too short \citep[][but see also \citealt{2020Kadam} for TI induced by water vapour opacity transitions]{2015Armitage}. A gas giant planet that migrated into the inner $\sim 0.1$~AU may modulate TI cycles, producing more promising light curves \citep{2004LodatoClarke}. However, hot Jupiters are only hosted by $\sim 0.5$\% of FGK stars \citep{2016SanterneMoutou}. More promisingly, the planet itself may be tidally disrupted at these radii, flooding the inner disc with `new matter', simultaneously producing outbursts with requisite properties \citep{2012NayakshinLodato} while avoiding the `too many hot Jupiters' conundrum. Planets made by the classical core accretion scenario are too dense for this to work, but planets made by the gravitational instability (GI) in the outer massive disc \citep{2005Rafikov} migrate inwards very rapidly and can indeed power FUor-like flares \citep{2005VorobyovBasu,2010VorobyovBasu,2010BoleyHayfield}.

On the other hand, protoplanetary discs may have a two-layer structure \citep{1996Gammie}, with a `dead zone' (DZ) in the inner $0.1 \leq R \leq$~a~few~AU. When a sufficient amount of material accumulates in the DZ to heat it up, magnetorotational instability (MRI) develops, producing bursts with properties similar to those observed \citep[][]{2001Armitage, 2009ZhuHartmann, 2010ZhuHartmann, 2020Kadam, 2020VorobyovKhaibrakhmanov}. 

Only recently have accretion bursts also been found in high-mass young stellar objects (HMYSOs) with $M_* \sim 10-20\msun$: S255IR NIRS 3 \citep{2017Caratti, 2020Uchiyama}, NGC 6334I MM1 \citep{2017Hunter, 2018MacLeod}, G358.93-0.03 MM1 \citep{2019Brogan, 2021Stecklum}, and G323.46-0.08 \citep{2019Proven-Adzri}. Physical conditions in the circumstellar discs of HMYSOs are very much different from those in their low-mass counterparts, providing an exciting opportunity to break the degeneracy between the existing FUor models.

\section{MRI activation scenario}\label{sec:MRI}

Our numerical approach is based on the 1D time-dependent viscous disc evolutionary code described in \citet{2012NayakshinLodato}, which optionally includes a migrating planet (not used in this section). The physical setup of our model follows \citet{2001Armitage} very closely. When thermal and/or cosmic ray ionisation is unable to keep the disc MRI-active, the local disc surface density, $\Sigma$, is the sum of the DZ component, $\Sigma_{\rm d} = \Sigma - \Sigma_{\rm a}$ (in this case, $\Sigma > \Sigma_{\rm a}$), and the two actively accreting layers, both with $\Sigma_{\rm a}/2$. The DZ is in the midplane and is sandwiched by the active layers. The viscosity parameters of the DZ and the active zones are $\alpha_{\rm d} = 10^{-4}$ and $\alpha_a\gg \alpha_{\rm d}$, respectively. The DZ becomes active if the midplane temperature exceeds the activation temperature, $T_{\rm a}$. The cosmic ray ionisation imposes $\min (\Sigma_a/2) =100$~g~cm$^{-2}$, which implies that the whole disc becomes active if its surface density drops below $200$~g~cm$^{-2}$ (even if $T < T_{\rm a}$).

\cite{2001Armitage} set $\alpha_{\rm a} = 10^{-2}$ and $T_{\rm a} =800$~K, following \cite{1996Gammie}. Such parameter choices led to  bursts that last $\gtrsim 10^4$~yrs for HMYSOs, much too long to be compatible with observed bursts,  which only last a few years (see Table~\ref{tab:1}). More recent studies of MRI-active discs \citep{2010ZhuHartmann, 2014BaeHartmann, 2020Kadam, 2020VorobyovKhaibrakhmanov} point towards higher $T_{\rm a}$, so we set $T_{\rm a} =1200$~K. Additionally, we set $\alpha_{\rm a} = 0.1$ (as suggested by the magnetohydrodynamics simulations of \citealt{2020Zhu}); both of these choices reduce the duration of the MRI bursts, making them more compatible with observations (cf. Sect.\  \ref{sec:comparison}).

In this letter we fix $M_*=15 \msun$. Our simulations started with a very low initial disc mass; the mass was then deposited from the envelope into the disc at a constant rate, $\dot M_{\rm ext}$, in a Gaussian ring with radial profile $\dot\Sigma_{\rm ext} = \dot\Sigma_0 \exp[-(R-R_0)^2/\sigma_e^2]$, where $R_0 = 10$~AU, $\sigma_e = 0.1 R_0$, and $\dot\Sigma_0$ is a normalisation constant. Here we present results for $\dot M_{\rm ext}$=10$^{-5}~M_{\odot}\rm{yr^{-1}}$, which we found to be near optimal for this scenario. Although observations and numerical simulations suggest higher rates of gas infall from the envelope in HMYSOs, $\dot M_{\rm ext} \gtrsim$ a few $\times 10^{-4}\msun$~yr$^{-1}$ \citep{2020LiuZinchenko, 2018MeyerKuiper}, we find that at $\dot M_{\rm ext} > 5\times10^{-5}~\msun$~yr$^{-1}$ the MRI activation bursts disappear altogether as mass transport through the disc reaches a steady state. \footnote{This finding alone challenges this scenario as a possible mechanism of HMYSO bursts.} Lower values for $\dot M_{\rm ext}$ did not alter the results significantly either, save for lengthening the quiescent time between the bursts. 

\begin{figure}
\begin{centering}
\includegraphics[width=1\columnwidth]{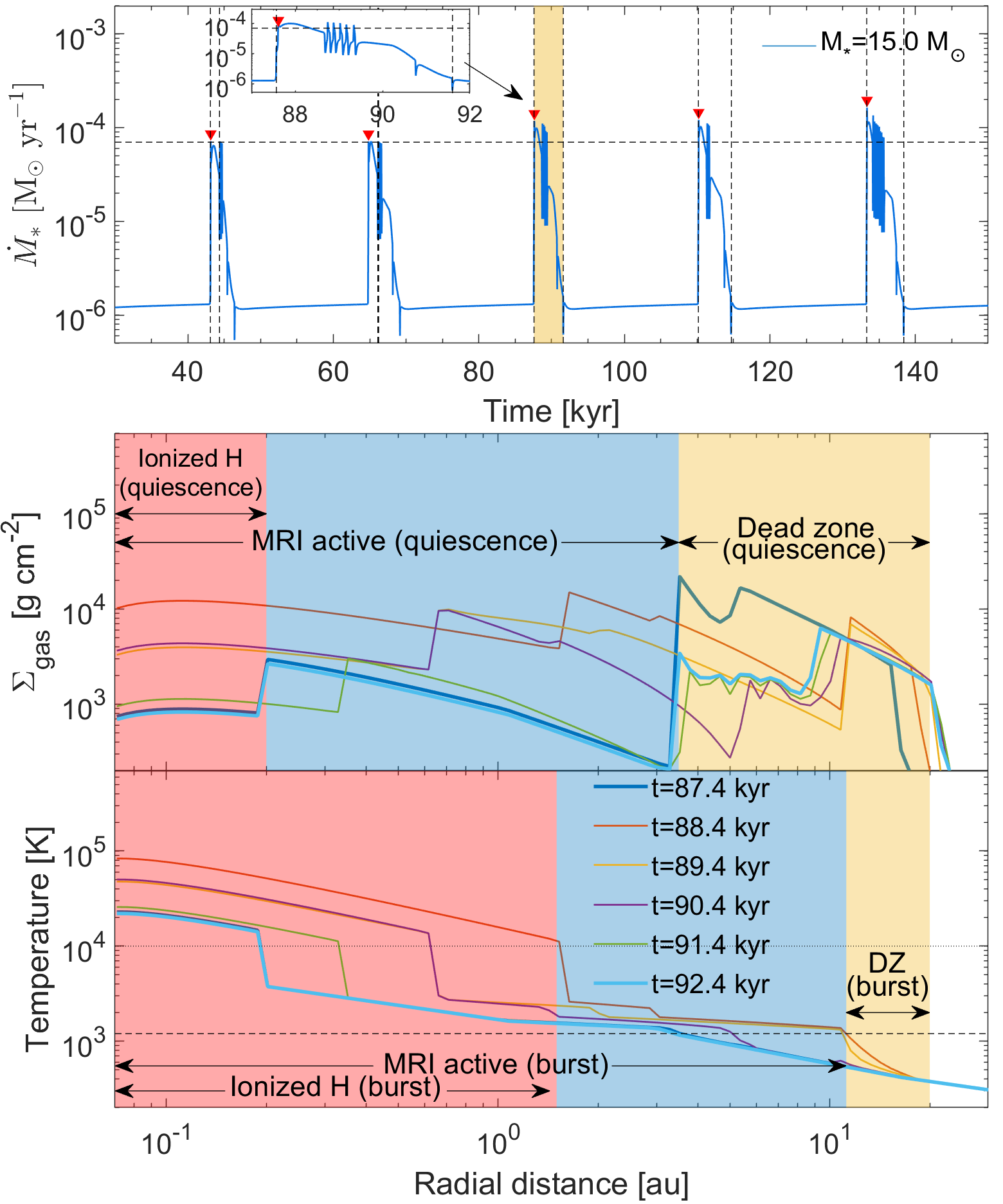}
\par\end{centering}
\caption{\label{fig:1} Accretion rate history and disc radial profiles in the MRI model. \textbf{Top:} Mass accretion rate as a function of time in the MRI model. 
The burst highlighted in yellow is also shown in the inset, zoomed in,  as well as in the central and bottom panels. \textbf{Middle and Bottom:} Temporal evolution of the radial distribution of surface density and temperature during the burst.
}
\end{figure}

Figure~\ref{fig:1} presents the temporal evolution of the stellar mass accretion rate, $\dot M_*(t)$, in the top panel, with the inset zooming in on one of the flares. The accretion rate during the outbursts is not quite as large as those observed ($\dot M_* \gtrsim 10^{-3} \msun$~yr$^{-1}$), but we were unable to reproduce that observational feature with this model.

The middle and the bottom panels of Fig. 1 show the disc surface density and the midplane temperature at various times during the burst highlighted in the top panel. The horizontal dashed line in the bottom panel shows the MRI activation temperature $T_{\rm a}=1200$~K, while the horizontal dotted line shows temperature $T_{\rm H}=10^4$~K, above which most of the hydrogen in the disc is ionised. The radial locations of the respective regions in the disc control the character of the outbursts that follow. Activation of MRI in the DZ is responsible for the bulk of the flare duration and properties, and the ionisation of hydrogen drives shorter TI outbursts on top of that (these shorter bursts are seen in the inset in the top panel).

The red shaded region in the middle panel marks the inner $R<0.2$~AU of the disc, where hydrogen is ionised even during quiescence. At the peak of the burst the ionised H zone grows to $R\sim 1.4$~AU (see the bottom panel). The MRI-active region during quiescence extends to R$\sim 3.5$~AU and to $R\sim 11$~AU during the burst. The DZ (yellow shaded zone) varies in extent from $3.5<R<20$~AU in quiescence to $11 < R < 20$~AU during the burst.

The MRI activation instability cycle proceeds as follows. During the quiescent time intervals, material accumulates in the DZ. When the inner boundary of the DZ heats up to $T>T_{\rm a}$, the viscosity increases as MRI is activated there. The resulting increased mass transfer heats up the disc, and the MRI-active zone grows to $R\sim 11$~AU. The material that accumulated in the DZ out to this radius then drains onto the star, powering a burst. During the flare, the extent of the ionised H zone may fluctuate, producing TI bursts \citep{1994BellLin} on top of the MRI burst.

By analysing the $\dot M_*(t)$ curve for the shaded burst in the top panel, we obtained MRI and TI burst durations of $t_{\rm dur}\sim 4500$ and 200 yrs, respectively (cf. Table 1). Appendix~B explains how burst durations are defined; it should be noted that there is a mild dependence here on the choices made. These durations should be of the order of the viscous time in the active state at the outer edge of the disc region participating in the outburst. For the MRI and TI bursts, we have 
\begin{equation}
    t_{\rm visc} = \frac{1}{\alpha\Omega_{\rm K}}\frac{R^2}{H^2} \approx 4000 \hbox{ yr}\frac{m_*^{1/2}}{\alpha_{-1}}  \left(\frac{R}{10\rm AU}\right)^{0.5} \left(\frac{1000\rm{K}}{T}\right)\;,
    \label{eq:tvisc0}
\end{equation}
where $\alpha_{-1} = \alpha/(0.1)$ and $m_* = M_*/(15 \msun)$. For TI bursts, the size of the active region is $R\approx 1.4$~AU, and Eq.~\ref{eq:tvisc0} with $T=10^4$~K yields $t_{\rm dur}\approx 182$~yr. These estimates are in excellent agreement with our $\dot M_*(t)$ curve analysis. Appendix~D shows how properties of TI bursts, studied independently of MRI activation bursts, vary with disc viscosity and the external accretion rate.

\section{Planet disruption scenario}\label{sec:PD_scenario}

\cite{2017MeyerVorobyov} found that circumstellar discs forming around HMYSOs produce GI-mediated accretion outbursts very similar in nature to those previously discovered for solar-mass YSOs \citep{2005VorobyovBasu,2010Vorobyov}. Further simulations by \cite{2019MeyerVorobyovElbakyan} and \cite{2020OlivaKuiper} show that GI-mediated bursts in HMYSOs can also be much brighter. Most recently, \cite{2021Meyer} performed a large parameter study of the model and showed that mass accreted per burst varies from $\sim 10\mj$ to $\sim 1\msun$. In Sect. \ref{sec:comparison} we see that, in the context of GI-mediated bursts, HMYSO bursts observed to date appear to require giant planet-mass objects. Henceforth, we considered an $M_{\rm p} = 5\mj$ mass planet migrating inwards towards the protostar. In Appendix~A we show that the formation of such objects by GI in the outer disc, and their arrival in the inner disc via migration, is indeed possible. Although our 1D simulation does not self-consistently model disc fragmentation, which requires multi-dimensional numerical methods, here we are able to follow planet migration and its tidal disruption in the very inner disc, $R\leq 1$~AU, where most of the observed accretion power is released.

The Hill radius of the planet, 
\begin{equation}
    r_{\rm H} = R \left(\frac{M_{\rm p}}{3M_*} \right)^{1/3} \approx 100 R_{\rm J} \left(\frac{R}{1\hbox{ AU}} \right) \left(\frac{M_{\rm p}}{5 M_{\rm J}} \right)^{1/3} \left(\frac{15 \msun}{M_*} \right)^{1/3}
,\end{equation}
shrinks rapidly as the planet migrates closer to the star. Planets larger than $\sim 10 R_{\rm J}$ will fill their Roche lobe at the tidal radius, $R_{\rm t}$ (where the planet radius $r_{\rm p}= r_{\rm H}$), before they reach the stellar surface. Following \cite{2012NayakshinLodato}, we assumed that the mass loss from the planet commences at $R= R_{\rm t}$  and that this mass is deposited into the disc via the Lagrange L1 point. Here we pick a time-independent mass-radius relation for the planet, 
\begin{equation}
r_{\rm p}(M_{\rm p}) = 30 R_J\left(\frac{M_{\rm p}}{\mj}\right)^{-1/3}
\label{R_vs_M}
,\end{equation}
as appropriate for a uniform composition polytropic planet with the mono-atomic ideal gas equation of state.  With this prescription,  $r_{\rm p} = 17 R_{\rm J}$ at $M_{\rm p} = 5\mj$ (i.e. before mass loss sets in). This is much larger than expected for a gas giant made by core accretion, but it is reasonable for a recently collapsed GI planet\footnote{GI protoplanets are born as very extended, $r_{\rm p} \gtrsim 1$~AU, low density clumps that are mainly composed of H$_2$ \citep[e.g.][]{2010Nayakshinc}. As they cool, they contract and can eventually collapse into denser configurations when H$_2$ molecules dissociate. These are the planets that we study here.} and is consistent with the values used in \cite{2012NayakshinLodato}. We note that $R_{\rm t} \approx 0.17$~AU for these parameters.

\begin{figure}
\begin{centering}
\includegraphics[width=1\columnwidth]{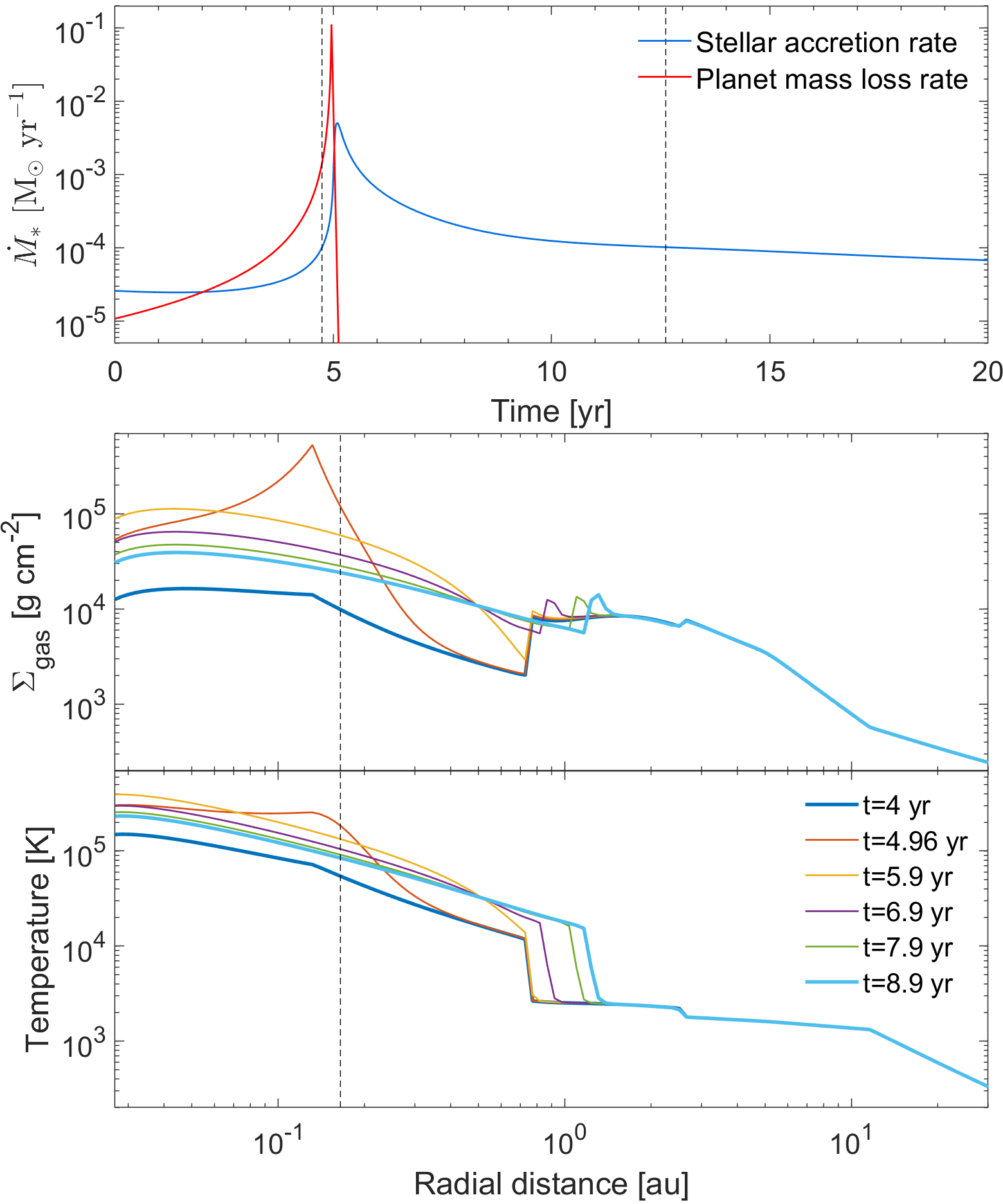}
\par\end{centering}
\caption{\label{fig:2}Accretion rate history and disc radial profiles in the PD model. \textbf{Top:} Stellar mass accretion rate history (blue line) during a short evolutionary time period when the planet is disrupted and the accretion burst is initiated. The red line shows the rate of mass loss by the planet. Vertical dashed lines mark the beginning and end of the burst.
\textbf{Middle and bottom:} Temporal evolution of the surface density and temperature distribution during the burst. The vertical dashed line shows the position of the planet at the moment of disruption.
}
\end{figure}

In Fig.~\ref{fig:2} we show just the short period of disc evolution during which the planet fills its Roche lobe and is tidally disrupted. The red curve shows the planet mass-loss rate, $\dot M_{\rm p}$, and the blue curve shows the accretion rate onto the star, $\dot M_*$. We note that the pre-outburst accretion rates are higher in this calculation than in Fig.~\ref{fig:1}. This was done for computational expedience as planets migrate more rapidly in high-$\dot M_*$ discs and are therefore disrupted more quickly. In Appendix~C we show that outburst profiles depend very weakly on pre-outburst $\dot M_*$. The middle and the bottom panels of Fig.~\ref{fig:2} show, respectively, the temporal evolution of the surface density and temperature distribution in the disc at times just before and during the planet disruption (PD) burst.
The vertical lines in the middle and bottom panels indicate the position of the planet at these times. The mass lost by the planet is deposited at the circularisation point of the material lost through the L1 point (i.e. just inwards of the planet orbital position). This corresponds to the highest peak in the $\Sigma$ curve in the middle panel. The peak in $\Sigma$ is then spread inwards and outwards by the viscous disc torques.  

The planet mass-loss rate is a strong function of $\Delta r=r_{\rm p}-r_{\rm H}$ and $ \dot M_{\rm p} \propto \Delta r^3$ for $\Delta r >0$ \citep{2012NayakshinLodato}. Since the planet migration is negligible over a time span of a few years, it can be shown that, once the Roche lobe overflow commences, the planet mass loss is a runaway process, scaling as
\begin{equation}
    \dot M_{\rm p}(M_{\rm p}) \propto \Delta r^3 \propto \left(\frac{1-x^2}{x}\right)^3\;,
    \label{eq:mdot}
\end{equation}
where $x = M_{\rm p}/M_{\rm p0}$, with $M_{\rm p0} = 5\mj$. This leads to a powerful accretion outburst in which the rise in the stellar accretion rate lags behind $\dot M_{\rm p}$ by just a fraction of a year. This lag is of the order of the local viscous time at $R\approx 0.1$~AU, $t_{\rm visc}\approx 1.3$~yr. Appendix~C shows that the outburst peak accretion rate is approximately $\sim M_{\rm p}/t_{\rm visc}$ and that the burst duration is of the order of $t_{\rm visc}$, largely independent of the pre-outburst disc state. The decay of the outburst occurs on a slightly longer timescale of a few years as some of the material is spread viscously outwards by a factor of a few in radius.

\section{Comparison of models with observations}\label{sec:comparison}

The recent observations of HMYSOs reveal significant luminosity brightenings, which are believed to be caused by accretion bursts on HMYSOs, from several sources: S255IR-NIRS3 \citep{2016Stecklum, 2017Caratti, 2018LiuSuZinchenko}, NGC6334I-MM1 \citep{2017Hunter, 2018Hunter, 2018Brogan}, and G358.93-0.03-MM1 \citep{2019Brogan, 2019MacLeod, 2021Stecklum}. In Table 1 we summarise the inferred properties of the bursts from these sources and compare them with the two representative calculations presented in the previous sections. 

\begin{table*}
\center
\caption{\label{tab:1} Main properties of model accretion outbursts compared to the HMYSOs observed so far. Burst durations in our models are calculated at vertical distances equal to 2\% and 10\% of the burst prominence.}
\begin{tabular}{cccccc}
\hline 
\hline 

Object/Model & Mass & Burst duration (2\%/10\%) & Rise time & $\dot{M}_{\rm peak}$ & Accreted mass \tabularnewline
& $M_{\odot}$ & yr & yr & $M_{\odot} \rm{yr}^{-1}$ & $M_J$  \tabularnewline
\hline 
Thermal Instability (TI) & 15 & 200/88 & 60 & 1.1$\times10^{-4}$ & 4.3 \tabularnewline
MRI activation & 15 & 4500/1150 & 50 & 1.2$\times10^{-4}$ & 103 \tabularnewline
Planet Disruption (PD) & 15 & 7.9/1.3 & 0.3 & 5$\times10^{-3}$ & 1.84 \tabularnewline \hdashline
S255IR NIRS 3$^\dagger$ & 20 & $2-2.5$ & 0.4 & 5$\times10^{-3}$ & 2 \tabularnewline
NGC 6334I MM1$^*$ & 20 & >6 & 0.62$^{+0.14}_{-0.14}$ & $10^{-3}$ & 0.325$^{+0.126}_{-0.094}$\tabularnewline
G358.93-0.03 MM1$^\ddag$ & 9.7$^{+0.3}_{-0.6}$ & 0.75 & 0.16$^{+0.01}_{-0.01}$  & $1.8^{+1.2}_{-1.1}\times10^{-3}$ & 0.566\tabularnewline

\hline 
\end{tabular}
\\
\textbf{Notes.}
$\dagger$\citet{2017Caratti}, \;
$^*$\citet{2017Hunter}, \;
$\ddag$\citet{2021Stecklum}.
\end{table*}

We find that MRI activation and TI scenarios yield outbursts that are too long in duration and too low in peak mass transfer rates. Ultimately, this is due to the large radial extent of the disc regions that are `on' during outbursts, $R\sim 10$~AU and $R\sim 1$~AU, respectively (cf. the lower panel of Fig. \ref{fig:1}). These bursts are excessively long compared with the burst durations observed thus far (from a few months to a few years). In addition, the rise times of TI and MRI outbursts are also too long compared with observations, for similar reasons. Although not presented here, we performed a wide sweep of the physically reasonable parameter space of these models, covering $10^{-2} \leq \alpha_{\rm a} \leq 0.3$, $800 \leq T_{\rm a} \leq 1500$ K, and external disc feeding rates from $10^{-6} \msun$~yr$^{-1}$ to $10^{-4} \msun$~yr$^{-1}$. All of these models resulted in bursts that are insufficiently luminous yet too long.

In contrast, PD bursts have temporal characteristics and a peak mass accretion rate that agree with those observed. 
Since the planet is disrupted at $R_{\rm t} \approx 0.17$~AU, the resulting bursts are much brighter and shorter than those from the MRI and TI scenario. In particular, for the case presented in Sect. \ref{sec:PD_scenario}, the peak accretion rate is $5\times10^{-3}~M_{\odot}\rm{yr}^{-1}$, which is more than an order of magnitude higher than the peak value for the TI and MRI bursts. The burst duration is about 8 years, and the rise time for the burst is about 4 months. 

Further support for the PD scenario comes from the mass budget of the observed bursts, that is, the mass accreted onto the star during the burst, shown as $\Delta M$ in the last column of Table 1. For the observed bursts, $\Delta M$ is of the order of a gas giant planet mass, similar 
to that obtained in the PD calculation. In contrast, in the MRI activation scenario, $\Delta M$ is is much higher, ranging from tens to hundreds of $\mj$, depending on the $\alpha_{\rm a}$ parameter in the active zone.

Finally, we note that while TI does not appear to be a convincing scenario for the powerful HMYSO bursts observed so far, this instability may be important for the lower amplitude variability of these sources. In Appendix D we show that HMYSO discs are susceptible to TI in a very wide range of conditions. This implies that some of the observed low-$\dot M$ sources may currently be in the quiescent part of the TI cycle, and that their time-averaged mass accretion rate is much higher.

\section{Discussion and outlook}\label{sec:discuss}

In this paper we studied how some of the  models for the classical FU Ori outbursts of LMYSOs scale to HMYSOs. We found that two well-known instabilities operating in the inner disc -- the TI and the MRI activation instability -- always produce outbursts that are both much too dim and too long (by hundreds to thousands of years) compared with the powerful, several-year-long outbursts observed from HMYSOs so far. Analysis of the disc viscous timescales shows that to explain such short outbursts, the disc must flare up in its very inner region,  $R\sim 0.1$~AU, and yet the TI- and MRI-active zones extend to much larger scales in discs of HMYSOs, to $R\sim 1$~AU and $R\sim 10$~AU, respectively. The PD scenario explored in Sect. \ref{sec:PD_scenario}, on the contrary, agrees well with the characteristics of the bursts observed from HMYSOs so far (mass accreted, mass accretion rate, burst rise time, and duration). 

Here we focused on high-mass protostars. A preliminary investigation of burst properties for a wide range of protostar masses (from $M_* = 0.2\msun$ to $M_*=20\msun$) shows that the burst properties vary in a systematic and very different fashion with $M_*$ for the MRI, TI and PD scenarios. This suggests that future observations of YSO accretion bursts across a wide range of $M_*$ is a new promising dimension with which to differentiate scenarios of episodic protostar accretion.

In the future we shall provide more detailed spectral and temporal predictions to test burst scenarios. A sudden injection of `new' mass into the inner disc sets PD bursts apart from those produced by any accretion disc instabilities. Such an injection results in an unexpected peak in the disc surface density profile and temperature (cf. the middle panel of Fig. \ref{fig:2}) at the PD location, which may carry observable imprints in the disc spectral energy distribution.

It is also important to investigate a broader range of properties for the objects migrating towards the star. Simulations \citep[e.g.][]{2021Meyer} predict that brown dwarfs and very young low-mass protostars could also be driven rapidly into the inner disc. While the mass budget of bursts observed from HMYSOs so far is in the planetary mass regime, it may be possible for a more massive object to lose only a part of its envelope if its mass-radius relation, $r(M)$, is different from the simple power law we used here (Eq. \ref{R_vs_M}).

Here, we studied an individual PD burst event. Simulations of gravitationally unstable discs show the formation of numerous gas clumps, even in solar-mass YSO discs \citep{2011ChaNayakshin, 2013VorobyovZakhozhay}. We may hence expect many PD events to occur in HMYSOs \citep{2018MeyerKuiper}. Burst statistics are, however, likely to be very different from those produced by MRI and TI instabilities. The arrival of a planet (or a more massive object) in the inner disc is likely to be a stochastic process. Therefore, we do not expect PD bursts to be periodic. Their individual properties may also vary significantly, depending on the object mass and radius (the latter controls the location where the disruption takes place in the disc). On the other hand, MRI bursts show periodicity with the quiescent phases a few times longer than the burst duration. The durations of TI bursts are comparable with those of the corresponding quiescent phases between them (see Appendix~D). Such a regularity for TI bursts can serve as a predictive tool for indicating the approximate start time of the next burst. It may also be used to try to trace back the preceding TI burst events. This may be useful, for example, in protostellar jet studies, where shock fronts in the jets are attributed to the burst events at the jet base \citep{2018VorobyovElbakyanPlunkett}.  Additionally, chemical and/or kinematic signatures of past outbursts in the outer disc or the surrounding envelope material may also tell us about burst periodicity and nature \citep{2019Wiebe, 2021Molyarova, 2021VorobyovElbakyanLiu} .

\section{Conclusions}

In this paper we have shown that accretion bursts recently observed in HMYSOs are unlikely to result from the disc hydrogen ionisation or the MRI activation instabilities that were previously proposed to power the FU Ori outbursts of LMYSOs. We also showed that the tidal disruption of young gas giant planets formed in the outer disc by the fragmentation of the gravitationally unstable disc explains observed burst characteristics well, provided that planetary radii are $R_{\rm } \gtrsim 10  R_{\rm J}$. Future modelling and observations of protostellar accretion bursts promise to be sensitive probes of both the inner disc structure and  the fragmentation of the outer disc due to GIs for low- and high-mass stars alike.

\begin{acknowledgements}
We  thank  the  anonymous  referee  for an insightful report, which helped to improve this paper. We thank Andrey Sobolev for useful discussions. V. E. and S. N. acknowledge the funding from the UK Science and Technologies Facilities Council, grant No. ST/S000453/1. This work made use of the DiRAC Data Intensive service at Leicester, operated by the University of Leicester IT Services, which forms part of the STFC DiRAC HPC Facility (www.dirac.ac.uk). V. E. also made use of funds from  the Ministry of Science and Higher Education of the Russian Federation (State assignment in the field of scientific activity, Southern Federal University, 2020).
E. I. Vorobyov acknowledges support from the Russian Science Foundation grant 18-12-00193.
A.C.G. has received funding from the European Research Council (ERC) under the European Union’s Horizon 2020 research and innovation programme (grant agreement No. 743029).
\end{acknowledgements}

\bibliographystyle{aa}
\bibliography{ref_base}

\newpage
\begin{appendix}
\label{sect:app}
\section{Planet migration in the disc of HMYSO}

Theoretical analysis of disc fragmentation conditions shows that the mass of the first fragments that form in self-gravitating discs -- which we call here the Toomre mass, $M_{\rm T}$ --  is largely independent of the mass of the central object \citep{2007Levin}. This is because these conditions \citep{2001Gammie} are very similar \citep{2010Nayakshinc}  to those setting the fragmentation opacity limit \citep{1976Rees}, hence leading to $M_{\rm T} \sim (1-10)\mj$ even around objects as massive as the $4\times 10^6\msun$ supermassive black hole in the centre of the Milky Way (see Fig. 1 in \citealt{2006Nayakshin} and Fig. 3 in \citealt{2007Levin}). However, this is not to say that we expect final fragment masses to be the same, independent of the environment in which they formed. A highly uncertain issue is how rapidly these initial gas condensations grow by the accretion of gas \citep{2017NayakshinDesert}. If gas accretion onto the fragments is efficient, then -- simply due to the mass budget of the disc -- the more massive the central object is, the more massive the end products of disc fragmentation can be.

\begin{figure}
\begin{centering}
\includegraphics[width=1\columnwidth]{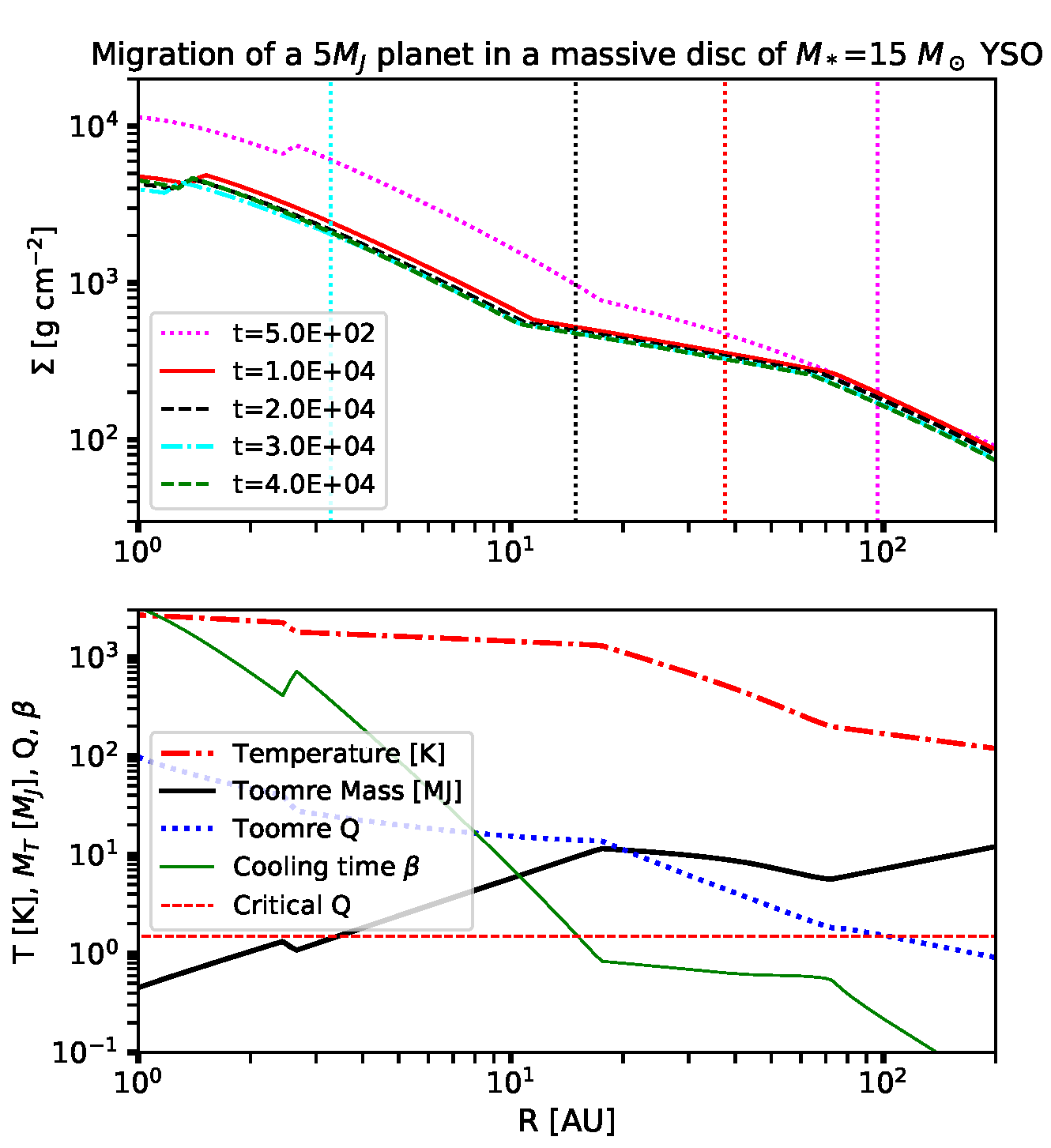}
\par\end{centering}
\caption{GI planet migration from its birth by disc fragmentation at 100 AU into the inner disc. \textbf{Top:} Disc surface density $\Sigma(R)$ at several different times, with the respective positions of the planet shown as vertical lines.  \textbf{Bottom:} Disc temperature, dimensionless cooling time, $\beta$, Toomre parameter, $Q$, and Toomre mass, $M_{\rm T}$ (initial fragment mass), at $t=2\times 10^4$ yr. The disc is GI unstable beyond $\sim 100$~AU and is capable of hatching gas giant planets.
}
\label{fig:OuterDisc}
\end{figure}

For the problem at hand,  Fig. \ref{fig:OuterDisc} shows the migration of a (constant mass) $M_{\rm p} = 5\mj$ planet injected into a disc with an initial surface density profile $\Sigma(R) \propto R^{-1}$ and with a total disc mass of $4\msun$ within $R=300$~AU rotating around a star with mass $M_*=15\msun$. Our model here and below in this section uses a fixed viscosity parameter $\alpha=0.1$ at all times (as in the active state in Sect. \ref{sec:MRI}). The top panel shows how $\Sigma(R)$ evolves, with vertical lines showing the corresponding radial position of the planet. We observe that it takes $\sim 0.04$~Myr for the planet to migrate into the inner few AU disc. We note that for HMYSOs, the planet migrates in the Type I regime (no gap opened) since the planet-to-star mass ratio is very small, $q = 3\times10^{-4}$.

The bottom panel of Fig. \ref{fig:OuterDisc} shows the disc midplane temperature, the dimensionless cooling time $\beta = t_{\rm cool}\Omega$, the Toomre parameter $Q$, and the initial fragment mass computed following \citet[][who took into account the corrections due to disc self-gravity and  the spiral arm geometry]{2010BoleyHayfield} as $M_{\rm T} = 1.6 M_* (H/R)^3$. Here, $t_{\rm cool}$ is the disc cooling time.
The dashed red horizontal line shows $Q_{\rm crit} = 1.5$. The disc is expected to fragment where $Q\leq Q_{\rm crit}$ and $\beta\leq 3$, which in this case corresponds to radii larger than $\approx 100$~AU. As suggested above, the Toomre mass is in the gas giant planet mass range at $R\sim 100$~AU, confirming that the fragmentation of this massive disc into planet-mass objects is possible. Qualitatively, the $M_{\rm T}$ does not differ greatly from that of $M_* \sim 1\msun$ protostars  because, while HMYSO discs are much hotter than LMYSO discs at the same $R$, they are geometrically thinner (i.e. they have smaller $H/R$) due to the much stronger gravity of their massive star.

This example calculation illustrates that the planet migration may indeed be sufficiently short to transport planet-mass objects into the very inner disc during a typical HMYSO evolutionary timescale, expected to be $\sim5\times10^4$~yr \citep{2021Meyer}.
Furthermore, for simplicity we assumed here that the protostar mass is $M_*=15\msun$ at the time of disc fragmentation. It is, however, quite likely that disc fragmentation begins when $M_*$ is far from reaching its final mass. In this case, the fragments can form closer in and migrate towards the star at faster rates while $M_*$ is moderately low. By the time the YSO reaches the observed $M_*\sim 10-20\msun$, these planetary mass fragments may have made most of their way into the inner 1 AU disc. Finally, Fig. \ref{fig:OuterDisc} assumes a constant planet mass. In a more realistic model, the mass of a planet may both increase and decrease during its migration, which would affect its migration speed. Stochastic interactions with spiral density arms, not taken into account in our azimuthally symmetric model, may shorten the migration time significantly for some of the fragments  \citep{2011BaruteauMeru}.

\section{Definition of burst duration}

We calculated the duration of each burst with respect to the prominence of the burst. The prominence of the burst measures how much the burst stands out due to its intrinsic height and its location relative to other peaks in the $\dot{M}_*$ versus time plot. To calculate the prominence, we found the minimum of the mass accretion curve on both sides of the peak. The higher of the two minima specifies the reference level. The height of the peak above this reference level is its prominence. We assumed the duration of the burst to be equal to the distance between the points where the descending total luminosity (on both sides from the peak) intercepts the horizontal burst duration line beneath the peak at a vertical distance equal to 2\% of the burst prominence.

\begin{figure}
\begin{centering}
\includegraphics[width=1\columnwidth]{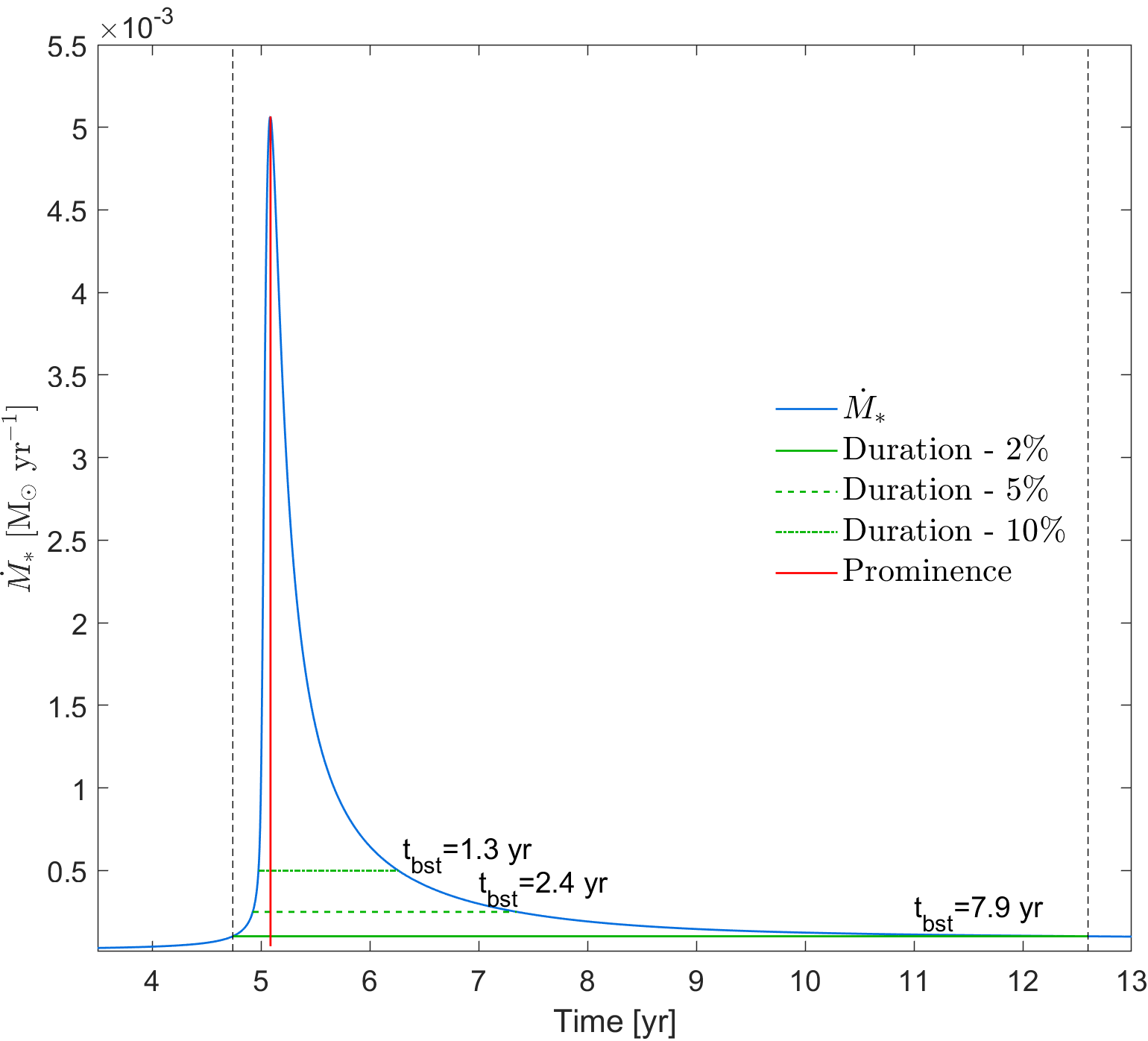}
\par\end{centering}
\caption{Stellar accretion rate history in the PD model during the burst event. The vertical red line shows the prominence of the burst, and the horizontal green lines show the durations of the burst measured at 2\%, 5\%, and 10\% of the prominence (from bottom to top).}
\label{fig:burst_dur}
\end{figure}

In Fig.~\ref{fig:burst_dur} we plot the short evolutionary period in the PD model during which the planet is disrupted and the accretion burst is initiated. The burst is similar to the one shown in the top panel of Fig.~\ref{fig:2}. The prominence of the burst is shown with the vertical red line, and the duration of the burst is shown with the horizontal solid green line. The vertical dashed lines mark the beginning and the end of burst. 

To show the importance of choosing the vertical distance from the bottom of the prominence at which the burst duration is measured, we plot in Fig.~\ref{fig:burst_dur} the burst durations measured at a vertical distance equal to 5\% (horizontal dashed green line) and 10\% (horizontal dash-dotted green line). The burst duration is 7.9 years for the vertical height of 2\% of the prominence and only 1.3 years for 10\% case.

\section{Planet disruption bursts depend weakly on disc properties}

\begin{figure}
\begin{centering}
\includegraphics[width=1\columnwidth]{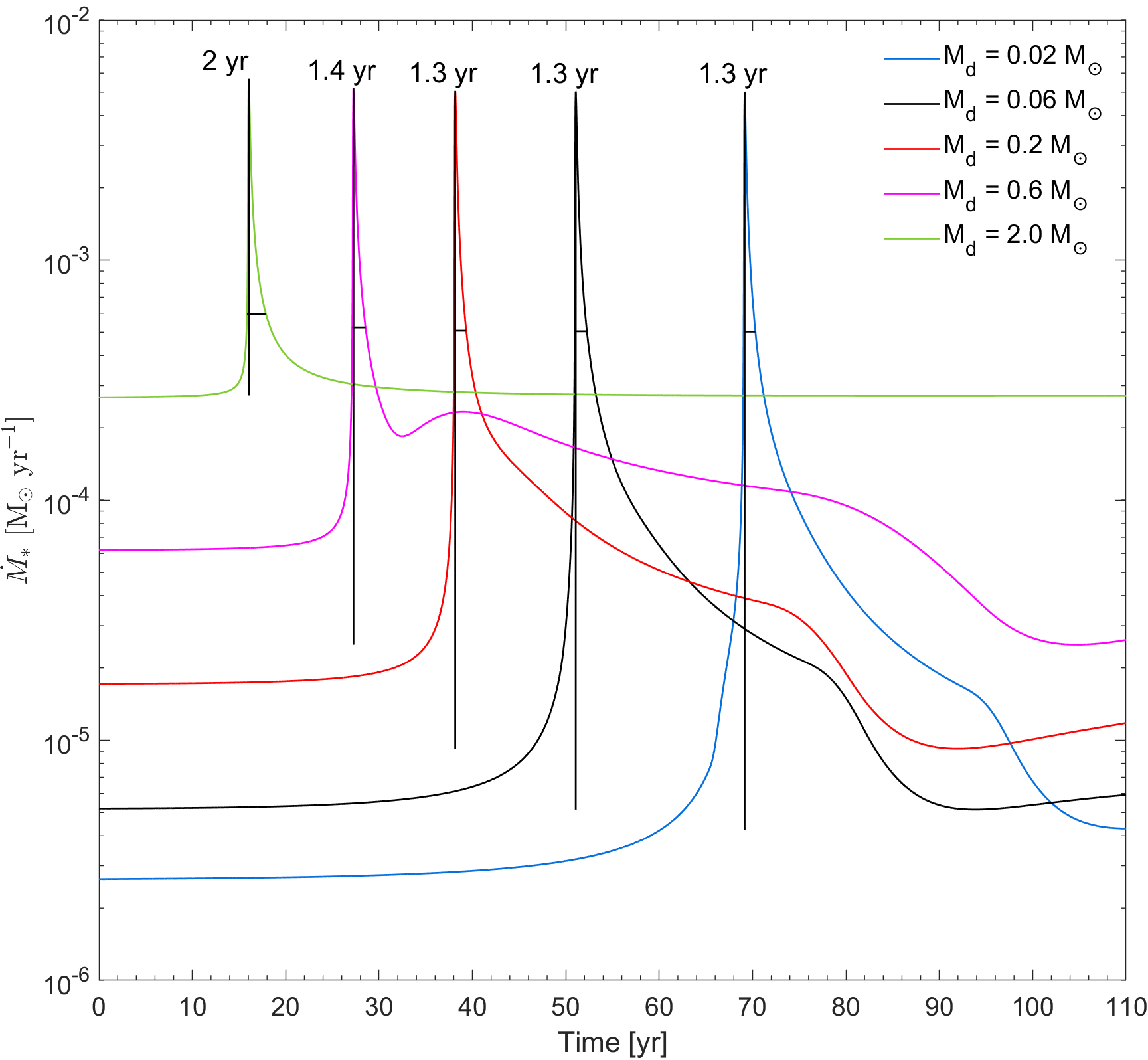}
\par\end{centering}
\caption{Accretion rate history during a PD burst in five models with different disc masses but with the same planet mass. Vertical black lines show the prominence of each burst, and the horizontal black lines indicate the duration of each burst measured at a vertical distance equal to 10\% of the burst prominence. The duration of the burst in years is shown at the top of each burst.}
\label{fig:arate_referee}
\end{figure}

Figure~\ref{fig:arate_referee} shows the mass accretion rate histories during a PD event in five different models that only differ from one another in terms of the initial disc mass ($M_{\rm d}$), and thus also the mass accretion rate in quiescence. The $M_{\rm d}$ values range from 0.02 to 2 $M_{\odot}$. Taken in its entire accretion rate range, the bursts appear somewhat different for these models. However, when analysing these curves we should distinguish the PD bursts from the disc variability due to the TI that is taking place even if the planet is not there. We find that TI bursts rarely rise above the accretion rate of a few$\times 10^{-4}\msun$~yr$^{-1}$. Therefore, if we focus on just the most luminous parts of the bursts in Fig. ~\ref{fig:arate_referee}, for example the  accretion rates exceeding 10\% of the peak (i.e. above $\sim 5\times10^{-4}~M_{\odot}$yr$^{-1}$), then we sample just the PD bursts. These bursts are remarkably similar to one another in terms of both the peak accretion rate and durations. This is because both of these factors are mainly determined by the viscous timescales of the disc at the radial distance of disruption, $t_{\rm visc}$. In particular, the PD burst duration is $\sim t_{\rm visc}$, whereas the maximum accretion rate is $\dot M_* \sim M_{\rm p}/t_{\rm visc}$.

By drawing these conclusions, we caution the reader that the situation may change depending on the mode of planet migration (gap opening, i.e. the Type II regime, versus no gap, i.e. Type I) and the internal structure of the planet as the interactions between the disc and the planet make this system highly non-linear \citep[cf.][]{2012NayakshinLodato}. These issues merit a deeper future study that would also cover a broader parameter space of the model.

\section{Dependence of thermal instability bursts on disc parameters}\label{sec:TI-appendix}

\begin{figure}
\begin{centering}
\includegraphics[width=1\columnwidth]{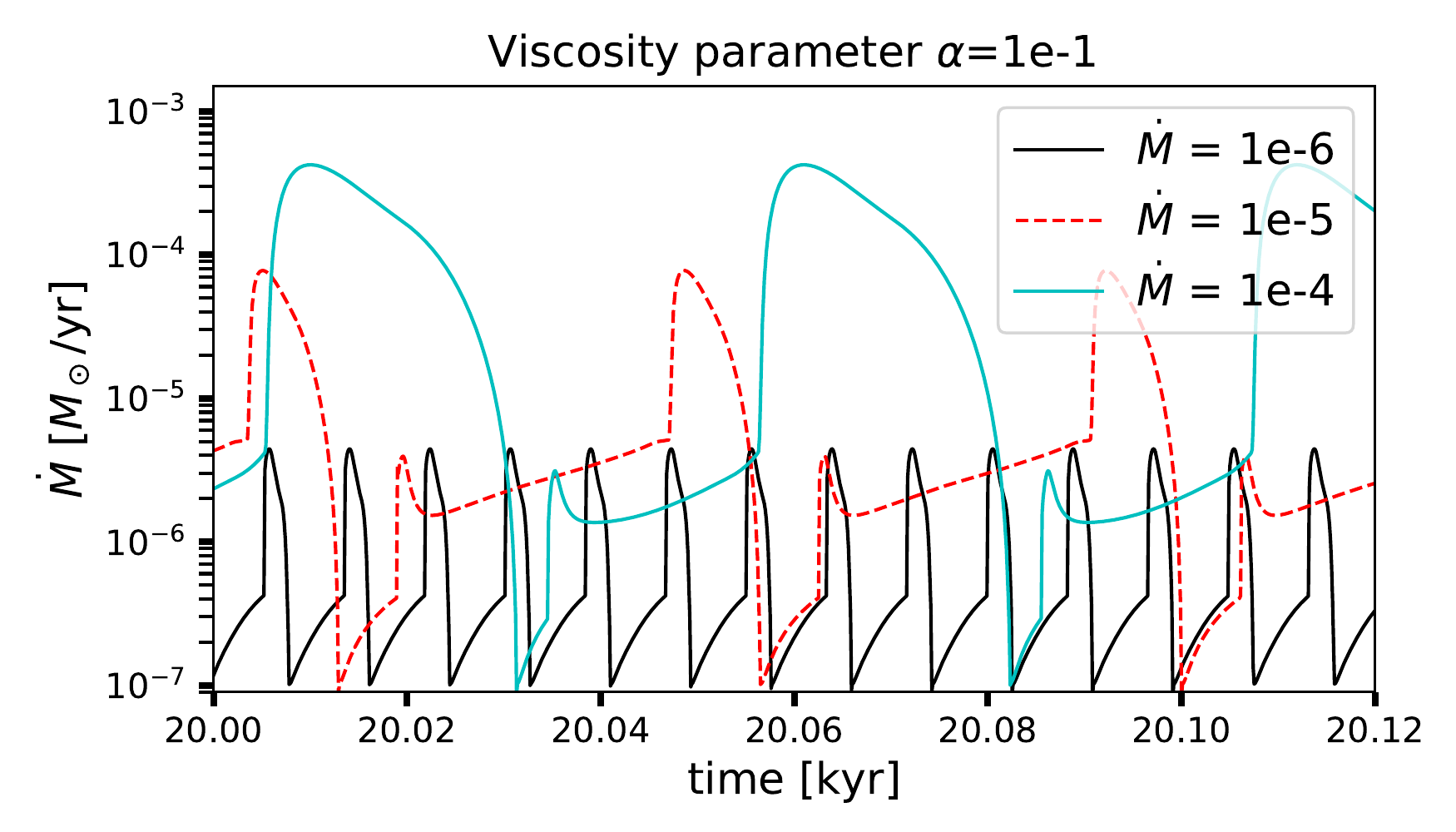}
\includegraphics[width=1\columnwidth]{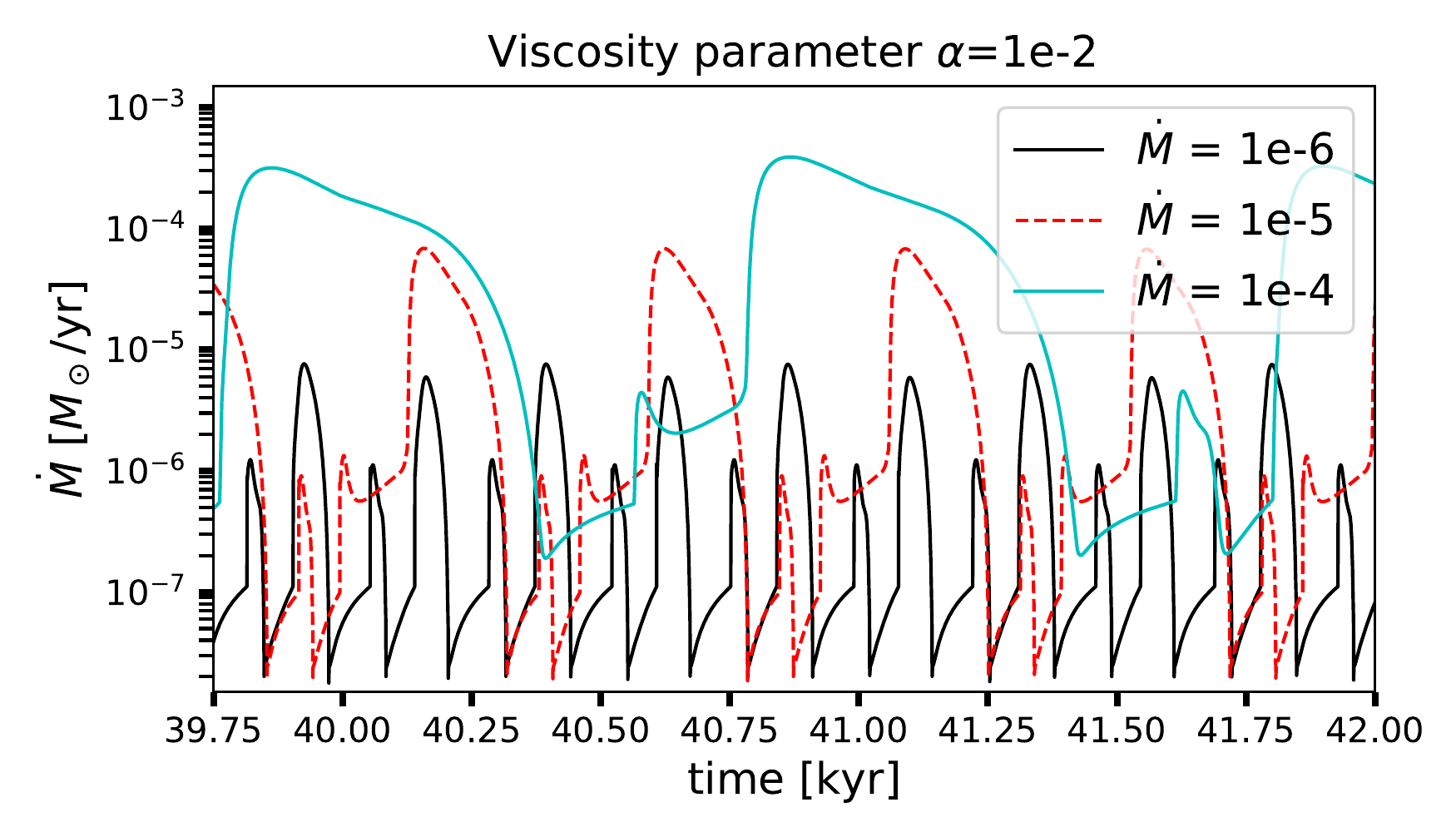}
\includegraphics[width=1\columnwidth]{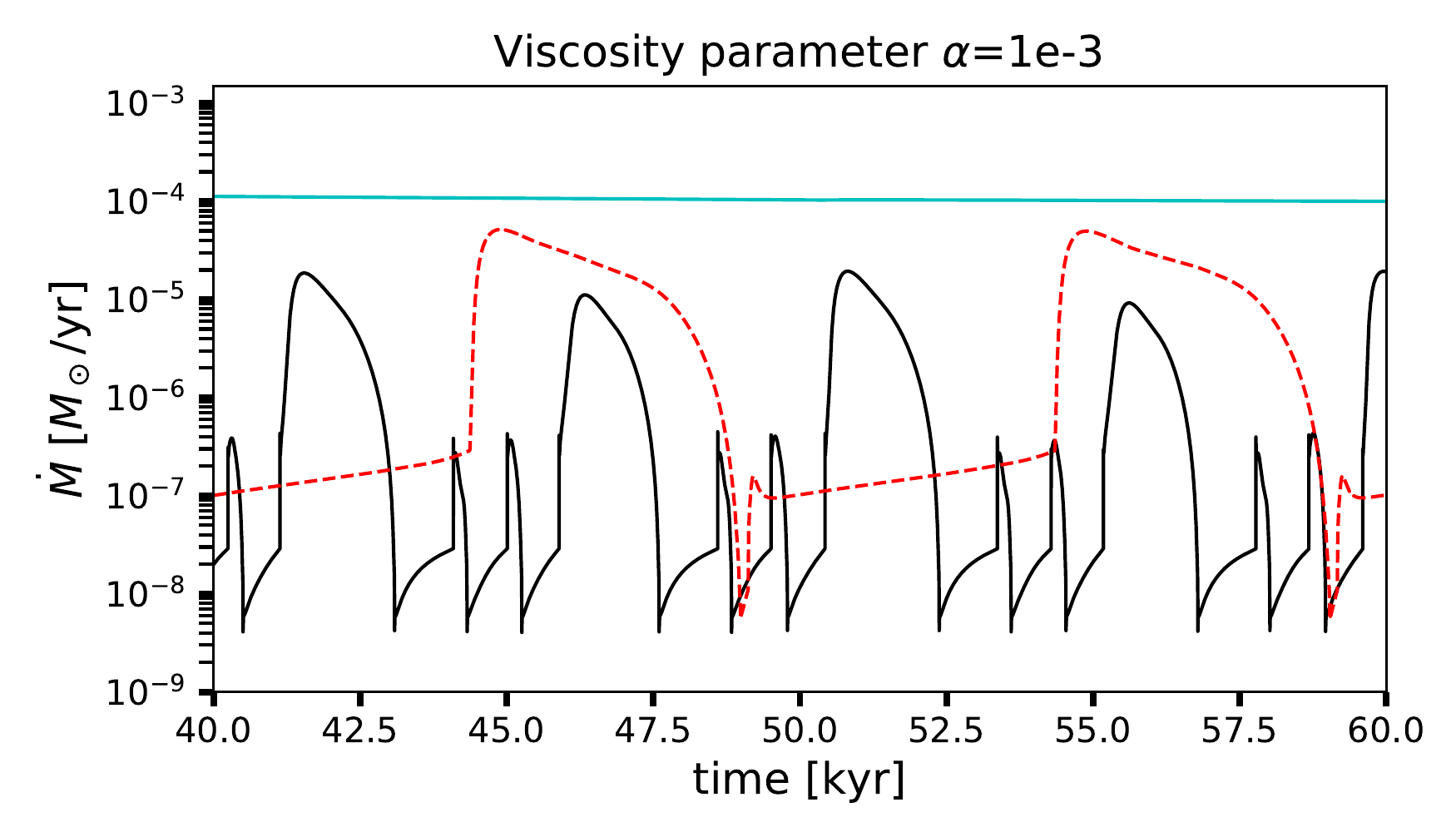}
\par\end{centering}
\caption{HMYSO mass accretion rates versus time for three different values of the (fixed) viscosity parameter, $\alpha$, as shown in the panel titles, for three different values of the external disc feeding rate, $\dot M_{\rm ext} = 10^{-6}, 10^{-5}, 10^{-4} \msun$~yr$^{-1}$, and $M_*=15~M_{\odot}$ .} 
\label{fig:alpha_1e-3}
\end{figure}

It is instructive to try to separate the TI  from the MRI activation instability. To this end, we present here an investigation of TI burst properties for varying external disc accretion rates and the disc viscosity parameter, $\alpha$. For the present calculation we set $\alpha$ to be the sum of a constant, which is independent of disc midplane temperature and $\Sigma$, plus the self-gravity term. We observe that, except for the single case of $\dot M_{\rm ext} = 10^{-4}\, \msun$~yr$^{-1}$ and $\alpha = 10^{-3}$, TI occurs in all of the parameter space studied. This indicates that TI is a very robust instability that occurs in the protoplanetary discs of HMYSOs and has to be taken into account if we are to interpret observations correctly.

Several other conclusions arise from Fig. \ref{fig:alpha_1e-3}. The peak accretion rates rarely exceed a few $\times 10^{-4} \msun$~yr$^{-1}$, which is an order of magnitude too low compared with the brightest of the HMYSO bursts observed so far. Additionally, the brightest parts of the bursts are rather periodic, and, as a rule, the duration of the `on' part of TI outbursts is about half of the period. This suggests that periodicity is a major property of TI outbursts. Such a periodicity is not expected from PD bursts; simulations show PD bursts to be chaotically spaced in the time domain \citep[e.g.][]{VorobyovBasu2015}.

\end{appendix}

\end{document}